# Rotating figures of equilibrium in General Relativity

T Papakostas

Department of Natural Resources and Environment Technological Educational Institute of Crete, 3 Romanou Street Chania 731-33 GREECE

E-mail:taxiar@chania.teicrete.gr

**Abstract.** A generalization of the notion of a surface of revolution in the spaces of General Relativity is presented .We apply this definition to the case of Carter's family [A] of spaces and we study Kerr metric with respect the existence of the above mentioned foliation.

## !.Introduction

The study of equilibrium structures of rotating fluid masses is a central problem in Newtonian Gravity and mostly in General Relativity where the solution of this problem constitutes an alternative to the existence of black holes.

After the investigations of Newton for the consequences of rotation of the earth to its shape, MacLaurin discovered in 1742, that spheroid configurations can be in equilibrium, when they rotate uniformly (MacLaurin spheroids) .About 100 years later, Jacobi discovered that non axial symmetric ellipsoids can emerge as equilibrium figures of uniformly rotating fluids (Jacobi ellipsoids 1834) .A general statement of the problem for homogeneous configurations has been formulated by Dirichlet in the framework of Lagrangian formalism, who solved it in 1875 for the case of an ellipsoid of revolution .Dedekind sticking on the equations obtained by Dirichlet,proved the existence of ellipsoids, congruent to those found by Jacobi,that are stationary in an inertial frame and maintain their ellipsoidal figure by the internal motions which prevail. In 1860 Riemann solved completely the problem posed by Dirichlet, under the assumptions of stationarity, uniform vorticity, with the velocity field linear in the coordinates, the resulting figures of equilibrium are known as the Riemann ellipsoids. The ellipsoids of Dedekind constitute the non axisymmetric branch of Riemann's sequence. The question of stability of the ellipsoid figures of equilibrium initiated by Poincare after his discovery of new pear-shaped configurations branched from the Jacobian sequence has been established rigorously in 1924 by H.Cartan. The definite consolidation and the interrelation of these classical results, as well as the stability of all sequences, has been achieved by

S.Chandrasekhar and N.Lebovitch in the sixties, see [1] for a complete review. The research within the framework of Newtonian Gravity is still in progress, many authors dropping the condition of homogeneity for the fluid, which seem to be inadequate for the description of real stars or galaxies.

The purpose of this brief survey of the Newtonian theory is to serve as a guide for the resolution of the analogue problem in the context of General Relativity that is the search for equilibrium fluid configurations satisfying Einstein's equations, bounded by a surface of zero pressure, matched across this surface to an asymptotically flat solution of Einstein's vacuum equations. In the context of this global model both the interior solution and the exterior solution are considered to be stationary and axially symmetric, see Geroh and Lindblom [2].

A similar attempt has been made by Krasinski [3] .In a remarquable paper he defined the notion of an ellipsoid of revolution in curved spaces and he tried to find perfect fluid solutions of Einstein's equations admitting a family of confocal ellipsoids as equipressure surfaces, unfortunately he have not found any new solution and his approach was not pursued further. After a long period of time Racz [4] considered stationary and axially symmetric vacuum spaces, possessing a one parameter congruence of confocal ellipsoids, generalizing the work of Krasinski. His purpose was to give a simple geometrical characterization for a class of Petrov type D solutions of Einstein's equations, namely the Kerr-NUT family. Recently Zsigrai [5], following the approach of Racz, generalized the definition of an ellipsoid in curved spaces and he presented as example stationary axially symmetric, perfect fluid space times with a so called confocal inside ellipsoidal symmetry.

We believe that this kind of approach can be very fruitful for the search of an interior configuration, (not necessarily corresponding to a perfect fluid), which could be matched to an exterior vacuum solution across an ellipsoidal surface of vanishing pressure. But we have an objection against the choice of an ellipsoidal bounding surface: the shape of the figures of equilibrium is not deduced by the field equations of the corresponding theory (Newtonian or General Relativistic) but it is imposed on them. It would be more reasonable to deduce the shape of the bounding surface from the equations governing the motion of the fluid; however this was never achieved until now.

The purpose of this paper is to propose a generalization for the bounding surface of the fluid; we don't look only for ellipsoidal surfaces, but we are searching for the most general surface compatible with the symmetries of our problem: stationarity and axial symmetry. Evidently these surfaces are the surfaces of revolution, which are obtained by revolving a plane curve C about a line L in its plane. The line L will be identified with the axis of rotation of the fluid configuration. Then the imposition of the Einstein's equations will permit to characterize the curve C and to study it using the well standard techniques of Differential Geometry. Obviously this family of surfaces includes as special case the ellipsoids of revolution, the study of non axially symmetric surfaces (triaxial ellipsoids) is excluded but this is not restrictive because in General Relativity the stationarity implies axial symmetry ,Lindblom [6],leaving the search for non axisymmetric configurations in the context of non stationary  spaces.

The main part of this paper is concerned with the construction of a frame work for considering surfaces of revolution in General Relativity, following the analogue approach of Krasinski for the ellipsoidal space-times. Then we begin to examine the applicability of the constructed frame work in the cases of stationary axially symmetric interior solutions of General Relativity. However as the Einstein's equations for the case of stationary and axially symmetric spaces are difficult to solve, we restrict our selves to a particular class of these spaces, the Carter's family [A] of spaces which contains many important solutions, see [7] for a complete presentation for the case of perfect fluid configurations.

In section 2 we present a treatment of stationary and axially symmetric spaces as well as the Carter's family [A] of solutions, we give also a brief description of the Newman-Penrose formalism in the context of complex vectorial formalism of Cahen,Debever,Defrise [8],[9],[10] . In section 3 we give the definition of a Riemannian 3-space foliated with surfaces of revolution and the definition of a curved 4-dimensional space whose quotient space with respect a class of commoving observers is the above mentioned 3-dimensional Riemannian space. In section 4 we apply these definitions to the case of Carter's class of solutions , we state the sufficient conditions for this type of foliation and we present a study of the Kerr metric in this context. Finally in section 5 we give the perspectives of our work and we propose the next problems to consider .

## 2.Stationary, axially symmetric spaces and Carter's family [A] of solutions

An important advance in the study of stationary and axially symmetric spaces in General Relativity was made by Papapetrou [11] when he showed, in the vacuum case that in any connected region containing the axis of symmetry, the orbits of the two parameter Abelian isometry group admit orthogonal two surfaces. The work of Papapetrou has been extended by Carter, who introduced the useful concepts of orthogonal transitivity and invertibility of the isometry group. We recall that an isometry group in a n-dimensional pseudo Riemannian space is said to be orthogonally transitive, if the p-dimensional orbits of the group are orthogonal to a family of (n-p)-dimensional surfaces. The group is said to be invertible if the isotropy subgroup contains an element of order two (an involution) which inverts the sense of p independent directions on the surface of transitivity at a point P but it leaves unaltered the directions orthogonal to the surface of transitivity at P. It was noticed by Kundt and Trumper [12] and by Carter [13] that Papapetrou theorem could be extended to the case when matter is present provided the Ricci tensor is invertible in the group.

**Hypothesis 2.1**. We suppose that the space admits an isometry Abelian group $G_2$ invertible, with non null surface of transitivity [14].

Hypothesis 2.1 implies that the metric tensor can be written in the following form:

$$ds^2 = g_{tt}dt^2 + 2g_{tz}dtdz + g_{zz}dz^2 - g_{xx}dx^2 - g_{yy}dy^2 \qquad (2.1)$$

The components of the metric tensor (2.1) $g_{ij}$ depend only on $x$, $y$ the group $G_2$ is generated by $\frac{\partial}{\partial t}$ (time-like Killing vector, implying the stationarity) and $\frac{\partial}{\partial z}$ (space-like Killing vector, implying the axial symmetry).The invertibility of the group is obvious for the fact that the transformation:

$$(t, z) \rightarrow (-t, -z) \tag{2.2}$$

is an isometry.The form (2.1) is identical with that used by Krasinski, but we use an alternative form appropriate for the calculations, the symmetric tetrad in which the Newman-Penrose (NP) coefficients are equal by pairs:

$$ds^2 = (Ldt + Mdz)^2 - (Ndt + Pdz)^2 - S^2 dx^2 - R^2 dy^2 \tag{2.3}$$

Where the functions L, M, N, P appeared in (2.3) are real and depend only on $x$, $y$.This form has been used by Carter [14] in the introduction of his spaces, by Debever [15], Carter and McLenaghan [16] and the author [17].

**Hypothesis 2.2.**The spaces (2.3) admit a second order Killing tensor, with two double non constant eigenvalues.

The spaces satisfying hypothesis 2.2 are Carter's family [A] of solutions, [17], [18] and the corresponding metric is:

$$ds^2 = (x^2 + y^2)\left\{\frac{E^2(y)}{(x^2+y^2)^2}(dt - x^2 dz)^2 - \frac{H^2(x)}{(x^2+y^2)^2}(dt + y^2 dz)^2 - \frac{x^2 dx^2}{F^2(x)} - \frac{y^2 dy^2}{G^2(y)}\right\} \tag{2.4}$$

The Carter's family [A] of solutions is characterized by the fact that the Hamilton-Jacobi (HJ) equation for the geodesics is solvable by separation of variables. The separation of variables takes place in a particular way and gives rise to a fourth constant of motion for the particle's orbits (the other three being the rest mass of the particle, the energy and the angular momentum about the symmetry axis).This quadratic in the velocity integral is due to the existence of the Killing tensor mentioned in hypothesis 2.2 and the relation between separability of the HJ equation and Killing tensors is studied in [19].The special kind of separation of variables imposed by Carter is translated at the level of the Killing tensor by the particular canonical form of hypothesis 2.2 and the Killing tensor can be written in the null tetrad of the NP formalism, as follows:

$$K_{ij} = \lambda_1(n_i l_j + l_i n_j) + \lambda_2(\bar{m}_i m_j + m_i \bar{m}_j) \tag{2.5}$$

With the appropriate choice of the null tetrad for the metric (2.4) we can calculate the corresponding spin coefficients which are equal by pairs:

$$\kappa = \nu = \sigma = \lambda = 0$$

$$\pi = \tau = \frac{\sqrt{2}}{2} \frac{H(x)}{(x^2+y^2)^{\frac{3}{2}}} \frac{G(y)}{E(y)} + i\frac{\sqrt{2}}{2} \frac{F(x)}{(x^2+y^2)^{\frac{3}{2}}}$$

$$\mu = \rho = -\frac{\sqrt{2}}{2} \frac{G(y)}{(x^2+y^2)^{\frac{3}{2}}} - i\frac{E(y)}{(x^2+y^2)^{\frac{3}{2}}} \frac{F(x)}{H(x)} \tag{2.6}$$

$$\gamma = \varepsilon = \frac{\sqrt{2}}{4} \frac{G(y)}{y(x^2+y^2)^{\frac{1}{2}}} \left[\frac{E_y}{E} - \frac{y}{(x^2+y^2)}\right] - i\frac{\sqrt{2}}{4} \frac{E(y)}{(x^2+y^2)^{\frac{3}{2}}} \frac{F(x)}{H(x)}$$

$$\beta = \alpha = \frac{\sqrt{2}}{4} \frac{H(x)}{(x^2+y^2)^{\frac{3}{2}}} \frac{G(y)}{E(y)} - i\frac{\sqrt{2}}{4} \frac{F(x)}{x(x^2+y^2)^{\frac{1}{2}}} \left[\frac{H_x}{H} - \frac{x}{x^2+y^2}\right]$$

Where $H_x = \frac{d}{dx}H(x)$ and $E_y = \frac{d}{dy}E(y)$.

The symmetric null tetrad in which the spin coefficients are calculated, in the complex vectorial formalism of Cahen, Debever, and Defrise is defined as follows:

$$\theta^1 = \frac{\sqrt{2}}{2}\left[\frac{E(y)}{(x^2+y^2)^{\frac{1}{2}}}(dt - x^2 dz) + (x^2+y^2)^{\frac{1}{2}}\frac{ydy}{G(y)}\right] = n_i dx^i$$

$$\theta^2 = \frac{\sqrt{2}}{2}\left[\frac{E(y)}{(x^2+y^2)^{\frac{1}{2}}}(dt - x^2 dz) - (x^2+y^2)^{\frac{1}{2}}\frac{ydy}{G(y)}\right] = l_i dx^i \tag{2.7}$$

$$\theta^3 = \overline{\theta^4} = \frac{\sqrt{2}}{2}\left[\frac{H(x)}{(x^2+y^2)^{\frac{1}{2}}}(dt + y^2 dz) + i(x^2+y^2)^{\frac{1}{2}}\frac{xdx}{F(x)}\right] = -\overline{m_i}dx^i$$

The metric is then:

$$ds^2 = 2(\theta^1\theta^2 - \theta^3\theta^4) \tag{2.8}$$

The covariant null vectors $l_i$, $n_i$ are real and the other two $\overline{m}_i$, $m_i$ are complex conjugate. A basis for the space of complex self dual two-forms is given by:

$$Z^1 = \theta^1 \wedge \theta^3 \quad Z^2 = \theta^1 \wedge \theta^2 - \theta^3 \wedge \theta^4 \quad Z^3 = \theta^4 \wedge \theta^2 \tag{2.9}$$

The components of the metric in this basis are:

$$\gamma^{\alpha\beta} = 4\left[\delta^\alpha{}_{(1}\delta^\beta{}_{3)} - \delta^\alpha{}_2\delta^\beta{}_2\right] \quad \alpha,\beta,\gamma = 1,2,3 \tag{2.10}$$

The complex connection one-forms $\sigma^\alpha{}_\beta$ are defined by:

$$dZ^\alpha = \sigma^\alpha{}_\beta \wedge Z^\beta \tag{2.11}$$

The vectorial connection one-form is defined by:

$$\sigma_\alpha = \frac{1}{8} e_{\alpha\beta\gamma} \sigma^\beta{}_\delta \tag{2.12}$$

Where $e_{\alpha\beta\gamma}$ are the three dimensional permutation symbols. The tetrad components $\sigma_{\alpha b}$ than the NP spin coefficients. The explicit correspondence is:

$$\sigma_1 = \kappa\theta^1 + \tau\theta^2 + \sigma\theta^3 + \rho\theta^4$$

$$\sigma_2 = \varepsilon\theta^1 + \gamma\theta^2 + \beta\theta^3 + \alpha\theta^4 \tag{2.13}$$

$$\sigma_3 = \pi\theta^1 + \nu\theta^2 + \mu\theta^3 + \lambda\theta^4$$

The complex curvature two-form $\Sigma^\alpha{}_\beta$ are defined by:

$$d\sigma^\alpha{}_\beta - \sigma^\alpha{}_\gamma \wedge \sigma^\gamma{}_\beta = \Sigma^\alpha{}_\beta \tag{2.15}$$

And the vectorial two-form by:

$$\Sigma_\alpha = \frac{1}{8} e_{\alpha\beta\gamma} \gamma^{\gamma\delta} \Sigma^\beta{}_\delta \qquad (2.16)$$

On expanding $\Sigma_\alpha$ in the basis $\left[ Z^\alpha, \overline{Z}^\alpha \right]$ one obtains:

$$\Sigma_\alpha = (C_{\alpha\beta} - \frac{1}{6} R\gamma_{\alpha\beta})Z^\beta + E_{\alpha\bar{\beta}} \overline{Z}^\beta \qquad (2.17)$$

Where the quantities $C_{\alpha\beta}$ and $E_{\alpha\bar{\beta}}$ are related to the NP curvature components $\Psi_A$, $\Phi_{AB}$ as follows:

$$C_{\alpha\beta} = \begin{pmatrix} \Psi_0 & \Psi_1 & \Psi_2 \\ \Psi_1 & \Psi_2 & \Psi_3 \\ \Psi_2 & \Psi_3 & \Psi_4 \end{pmatrix} \qquad E_{\alpha\bar{\beta}} = \begin{pmatrix} \Phi_{00} & \Phi_{01} & \Phi_{02} \\ \Phi_{10} & \Phi_{11} & \Phi_{12} \\ \Phi_{20} & \Phi_{21} & \Phi_{22} \end{pmatrix} \qquad (2.18)$$

And

$$6\Lambda = -\frac{R}{4}$$

R being the scalar curvature.

The null tetrad components of the Weyl tensor and the Ricci traceless tensor in the tetrad (2.7) satisfy the following relations:

$$\Psi_0 = \Psi_4 \qquad \Psi_1 = \Psi_3 \qquad \Psi_2 \neq 0$$

$$\Phi_{00} = \Phi_{22} \qquad \Phi_{02} = \Phi_{20} \qquad \Phi_{01} = \Phi_{21}$$

In the case of vacuum Einstein's equations, $E_{\alpha\bar{\beta}} = 0$ we obtain the following expressions for the functions of the metric (2.4):

$$G^2(y) = y^2 E^2(y) \qquad F^2(x) = x^2 H^2(x)$$

(2.19a)

$$E^2(y) = \frac{1}{2}by^2 + dy + p \qquad H^2(x) = -\frac{1}{2}bx^2 + cx + p$$

Where b, c, d, p are constants of integration. The Kerr metric is obtained if we set :

$$b = 2 \quad d = -2m \quad c = 0 \quad p = a^2 \qquad (2.19b)$$

And we make the following coordinate change:

$$y = r \qquad x = a\cos\theta \qquad (2.19c)$$

m is the mass, $a$ the angular momentum per unit mass and $r, \theta$ the Boyer-Lindquist coordinates. If we redefine the remaining two coordinates t and z as follows:

$$dz = a d\varphi \qquad dt = d\tilde{t} + d\varphi \qquad (2.19d)$$

we get the Kerr metric in the canonical form of the Boyer-Lindquist:

$$ds^2 = \frac{\Delta}{\rho^2}\left[dt - a\sin^2\theta d\varphi\right]^2 - \frac{\sin^2\theta}{\rho^2}\left[-adt + (r^2 + a^2)d\varphi\right]^2 - \frac{\rho^2}{\Delta}dr^2 - \rho^2 d\theta^2 \qquad (2.20)$$

Where $\Delta = r^2 - 2mr + a^2$ and $\rho^2 = r^2 + a^2\cos^2\theta$.

The Einstein's equations in the presence of a perfect fluid reduce to two equations [20]:

$$\Phi_{00}\Phi_{02} = \Phi_{01}^2 \qquad (2.21)$$

$$\Phi_{00} + \Phi_{02} = 2\Phi_{11} \qquad (2.22)$$

These equations imply that the energy-momentum tensor admits one simple eigenvalue (the energy-mass density) and a triple eigenvalue (the isotropic pressure). The solution of (2.21) permit to find G(y) and F(x) as functions of E(y) and H(x) and the coordinates x and y :

$$W(y) = \frac{G^2(y)}{E^2(y)} = k_4 y^4 + k_2 y^2 - k_0 \qquad Z(x) = \frac{F^2(x)}{H^2(x)} = -k_4 x^4 + k_2 + k_0 \qquad (2.23)$$

Where $k_0, k_2, k_4$ are constants of integration. Now if we put :

$$k_4 = -q^2 \qquad l^2 = k_2^2 - 4k_0 q^2$$

And we define a set of new coordinates $\xi$ and $\varsigma$ (Wahlquist coordinates) by the relations:

$$x^2 = l\xi^2 - \frac{k_2 - l}{2q^2} \qquad y^2 = l\varsigma^2 + \frac{k_2 - l}{2q^2}$$

We can finally solve equation (2.22) to obtain a generalization of the Wahlquist solution, [20].

### 3. Stationary axial symmetric spaces admitting a congruence of surfaces of revolution

We consider three dimensional Euclidean space foliated with surfaces of revolution with centre O and common axis of rotation L. Each surface is obtained by revolving a plane curve C about the axis of rotation L, this axis coincides with the axis of rotation of the fluid configuration. In a Cartesian coordinate system $(x_1, x_2, x_3)$, the curve C lies in the plane $Ox_1 x_3$ and $Ox_3$ is the axis of revolution. In this coordinate system, a parametric representation of the curve C is defined as follows:

$$x_1 = h_1(t) \qquad x_3 = h_2(t) \qquad t_1 \leq t \leq t_2 \tag{3.1}$$

Where t is the corresponding parameter. The elimination of the parameter between equations (3.1) permits to write the Cartesian equation of the curve:

$$x_1 = f(x_3) \tag{3.2}$$

The Cartesian equation of the surface of revolution will be:

$$f(x_3) = \sqrt{x_1^2 + x_2^2} \tag{3.3}$$

A parametric representation of this surface is given by the following relations:

$$x_1 = h_1(t) \cos \Phi$$

$$x_2 = h_1(t) \sin \Phi \tag{3.4}$$

$$x_3 = h_2(t)$$

Now we can define a system of coordinates adapted to the surface of revolution. From (3.4) it is clear that one of the coordinates will be $\Phi$, the azimuthal angle measured around the symmetry axis. The other two coordinates will be $r$ and $\theta$, some kind of generalized spherical coordinates. This system $(r, \theta, \Phi)$ suggests that the parameter t in (3.4) can be chosen to be the angle $\theta$, so the equation of the surface will be:

$$r = r(\theta) \tag{3.5}$$

And the parametric representation will have the form:

$$x_1 = h_1(r, \theta) \cos \Phi$$

$$x_2 = h_1(r, \theta) \sin \Phi \tag{3.6}$$

$$x_3 = h_2(r, \theta)$$

The corresponding metric in the 3 - dimensional Euclidean space assumes the expression:

$$ds_3^2 = (h_{1r}^2 + h_{2r}^2)dr^2 + (h_{1\theta}^2 + h_{2\theta}^2)d\theta^2 + h_1^2 d\Phi^2 + 2(h_{1r}h_{1\theta} + h_{2r}h_{2\theta})drd\theta \tag{3.7}$$

Where $h_{1r} = \dfrac{\partial h_1}{\partial r}$ , $h_{1\theta} = \dfrac{\partial h_1}{\partial \theta}$ etc.

The new coordinates are orthogonal only when :

$$h_{1r}h_{1\theta} + h_{2r}h_{2\theta} = 0 \tag{3.8}$$

We restrict our research to orthogonal coordinates for reasons of simplicity and geometric intuition. In fact in the case of ellipsoids of revolution, the condition (3.8) implies that the ellipsoids are confocal, we think that the surfaces of revolution have to share an analogue property. Then the metric of a surface of revolution reduces to:

$$ds_3^2 = (h_{1r}^2 + h_{2r}^2)dr^2 + (h_{1\theta}^2 + h_{2\theta}^2)d\theta^2 + h_1^2 d\Phi^2 \tag{3.9}$$

**Examples of surfaces of revolution.**

## A. Ellipsoids of revolution

The corresponding parametric representation will be:

$$h_1(r,\theta) = g(r)\sin\theta \qquad (3.10)$$

$$h_2(r,\theta) = r\cos\theta$$

The condition (3.7) reduces to :

$$gg_r - r = 0 \qquad (3.11)$$

The integration of this equation yields:

$$g^2 = r^2 \pm l^2 \qquad (3.12)$$

The equation of the surface (3.5) is simply $r$=constant and the Cartesian equation of the congruence of the ellipsoids of revolution is:

$$\frac{x_1^2 + x_2^2}{r^2 \pm l^2} + \frac{x_3^2}{r^2} = 1 \qquad (3.13)$$

It is important to note that the constant of integration $l$ is the measure of the ellipticity and the $\pm$ sign defines the shape of the ellipsoids (oblate for +, prolate for -).The corresponding metric is that obtained by Krasinski:

$$ds_3^2 = \frac{\left[r^2 \pm l^2 \cos^2\theta\right]}{(r^2 \pm l^2)} dr^2 + \left[r^2 \pm l^2 \cos^2\theta\right] d\theta^2 + (r^2 \pm l^2) d\Phi^2 \qquad (3.14)$$

## B. Tori of revolution

The parametric representation of a torus of revolution obtained by revolving an ellipse with centre situated at a distance R from the origin of the coordinate system, around the axis $Ox_3$, will be:

$$h_1(r,\theta) = g_2(r) + g_1(r)\sin\theta$$

$$(3.15)$$

$$h_2(r,\theta) = r\cos\theta$$

The integration of the corresponding equation (3.8) yields:

$$g_2 = R$$
And
$$g_1 = r^2 \pm l^2 \qquad (3.16)$$

The metric is written then as follows:

$$ds_3^2 = \frac{\left[r^2 \pm l^2 \cos^2\theta\right]}{(r^2 \pm l^2)}dr^2 + \left[r^2 \pm l^2 \cos^2\theta\right]d\theta^2 + \left[R + \sqrt{r^2 \pm l^2}\sin\theta\right]^2 d\Phi^2 \qquad (3.17)$$

The equation of the surface is just $r = $ constant.

The next step following the procedure of Krasinski is to generalize the metric (3.9) to that of a three dimensional curved space by adding an arbitrary function of $r$ and $\theta$ to the coefficient of $dr^2$, in such a way, that if the function is constant then we recover the corresponding Euclidean space (3.9). Also we are going to introduce a new set of coordinates $x$ and $y$ instead of $r$ and $\theta$, which is of great utility for the integration of Einstein's equations, the relation of the two coordinate systems is given by (2.19c). Then we can give the following definition:

**Definition 3.1**

The metric of Riemannian three dimensional space foliated with surfaces of revolution in an orthogonal system of coordinates is given by:

$$ds_3^2 = \frac{(h_{1y}^2 + h_{2y}^2)}{f^2(x,y)}dy^2 + (h_{1x}^2 + h_{2x}^2)dx^2 + h_1^2 d\Phi^2 \qquad (3.18)$$

And the functions $h_1(x, y)$ and $h_2(x, y)$ have to satisfy the orthogonality condition:

$$h_{1x}h_{1y} + h_{2x}h_{2y} = 0 \qquad (3.19)$$

In General Relativity the shape of a surface depends on the class of observers performing its description, taking in account the deformations due to relative motions. Obviously the equipressure surfaces of a fluid will be characterized as surfaces of revolution only for a particular class of observers. We consider that the motion of the commoving with the fluid observers have to be invariant under the action of the symmetry group, hence their velocity will be a linear combination of the Killing Vector fields $\frac{\partial}{\partial t}$ and $\frac{\partial}{\partial z}$.

**Definition 3.2**

The class of the commoving with the fluid observers has a velocity field of the following type:

$$\mathbf{u} = U(x,y)\frac{\partial}{\partial t} + V(x,y)\frac{\partial}{\partial z} \tag{3.20}$$

And since $g_{ij}u^i u^j = 1$ then:

$$g_{tt}U^2 + 2g_{tz}UV + g_{zz}V^2 = 1 \tag{3.21}$$

The next step is to consider when a 4 dimensional space-time is foliated with surfaces of revolution:

**Definition 3.3**

A 4 dimensional space-time is said to be foliated with surfaces of revolution if there exists a class of commoving with the fluid observers (definition 3.2), for which the union of their local spaces (their quotient space with respect (3.20)) is the spaces of definition (3.1).

$$ds_3^2 = \frac{(h_{1y}^2 + h_{2y}^2)}{f^2(x,y)}dy^2 + (h_{1x}^2 + h_{2x}^2)dx^2 + h_1^2 d\Phi^2 = (g_{ij} - u_i u_j)dx^i dx^j \tag{3.22}$$

And $$h_{1x}h_{1y} + h_{2x}h_{2y} = 0 \tag{3.23}$$

Then we can prove the following theorems:

**Theorem I**

The quotient space of the commoving with the fluid observers (Definition 3.2) is identified with a 3- dimensional Riemannian space foliated with surfaces of revolution (Definition 3.3) if the following conditions are satisfied:

$$V^2(g_{t\phi}^2 - g_{tt}g_{\phi\phi}) = h_1^2 \tag{3.24}$$

$$\frac{(h_{1y}^2 + h_{2y}^2)}{f^2} = g_{yy} \tag{3.25}$$

$$h_{1x}^2 + h_{2x}^2 = g_{xx} \tag{3.26}$$

This theorem resumes the projection $g_{ij} \to \gamma_{ij}$ where $\gamma_{ij}$ is the metric of the quotient space with respect the velocity field of the commoving observers (Definition 3.2):

$$\gamma_{ij} = g_{ij} - u_i u_j \tag{3.27}$$

The metric coefficients $g_{ij}$ are those of stationary and axially symmetric spaces (2.1) and the proof of the theorem is obtained by the identification (3.22).

The inverse transition $\gamma_{ij} \to g_{ij}$ which is necessarily not unique can be summarized by the following theorem:

**Theorem II**

The metric of a stationary and axially symmetric space-time $g_{ij}$, satisfying Definition 3.3 can be put in the following form:

$$g_{ij}dx^i dx^j = \left[\frac{(1-KV)}{U}dt + Kdz\right]^2 - h_1^2(dt - \frac{U}{V}dz)^2 - \frac{(h_{1y}^2 + h_{2y}^2)}{f^2}dy^2 - (h_{1x}^2 + h_{2x}^2)dx^2 \tag{3.28}$$

Where U and V are the functions defining the velocity field of the commoving with the fluid observers, K is an arbitrary function of x and y and the functions $h_1$ and $h_2$ have to satisfy (3.23).This expression is the generalization of formula (3.14) in the paper of Krasinski [3].The proof is based again on the identification (3.22), where the metric $g_{ij}$ is given by (2.1) and $u_i$ are obtained by (3.20) and (3.21).

**4.The case of Carter's class [A] of spaces**

We are going to consider now a special family of the stationary and axially symmetric spaces of General Relativity which includes many interesting solutions [10],[7] : Carter's family [A] of spaces. If we impose that the metric coefficients $g_{ij}$ of Theorem II are identified with those of metric (2.4) we can prove the following theorem:

**Theorem III**

The Carter's family of spaces [A] is said to be foliated with surfaces of revolution in the sense of Definition 3.3 if :

A. The velocity field of the commoving with the fluid observers of Definition 3.2 are completely defined :

$$V = \pm \frac{h_1}{EH} \tag{4.1}$$

$$U = \frac{1}{(E^2 - H^2)} \left\{ (x^2 E^2 + y^2 H^2)V + \left[ (E^2 - H^2)(x^2 + y^2) + h_1^2(x^2 + y^2)^2 \right]^{\frac{1}{2}} \right\} \tag{4.2}$$

B. Moreover:

$$\frac{(h_{1y}^2 + h_{2y}^2)}{f^2} = \frac{(x^2 + y^2)y^2}{G^2} \tag{4.3}$$

$$h_{1x}^2 + h_{2x}^2 = \frac{(x^2 + y^2)x^2}{F^2} \tag{4.4}$$

$$h_{1x}h_{1y} + h_{2x}h_{2y} = 0 \tag{4.5}$$

The proof is based on the replacement of the metric coefficients $g_{ij}$ in the theorems I and II by those of Carter's class [A] of solutions given by (2.4).The velocity field of the commoving with the fluid observer can be written in the tetrad (2.7):

$$u_i = \frac{\sqrt{2}}{2} \left\{ \Pi_1(n_i + l_i) + \Pi_2(m_i + \overline{m_i}) \right\} \tag{4.6}$$

Where $\quad \Pi_1 = \dfrac{E(U - x^2 V)}{(x^2 + y^2)^{\frac{1}{2}}} \quad$ and $\quad \Pi_2 = \dfrac{H(U + y^2 V)}{(x^2 + y^2)^{\frac{1}{2}}} \tag{4.7}$

The system of equations (4.3)-(4.5) is integrable and as a consequence we have to prove the following theorem:

**Theorem IV**

The system of equations (4.3)-(4.5), with unknown functions $h_1, h_2$, is integrable if the following relations hold:

A. The function $f(x, y)$ has to be of the form:

$$f(x, y) = \frac{f_1(y)}{f_2(x)} \tag{4.8}$$

B. The functions $G, F, f_1, f_2$ are related as follows:

$$u^2 = \frac{G^2}{f_1^2} = p_4 y^4 + p_2 y^2 - p_0 \tag{4.9}$$

$$v^2 = \frac{F^2}{f_2^2} = -p_4 x^4 + p_2 x^2 + p_0 \tag{4.10}$$

$$f_2 = \left\{ 2\sqrt{-p_4}\sqrt{-p_4 x^4 + p_2 x^2 + p_0} - 2p_4 x^2 + p_2 \right\}^{\frac{k}{2\sqrt{-p_4}}} \tag{4.11}$$

Where $k, p_0, p_2, p_4$ are constants of integration and $p_4$ has to be negative.

The proof of this theorem is based on the fact that the system of equations (4.3)-(4.5) can be solved by splitting it to two systems of P.D.E's:

$$h_{1x} = e_1 e_2 \frac{x(x^2 + y^2)^{\frac{1}{2}}}{F} \varepsilon$$

$$\tag{4.12}$$

$$h_{1y} = e_3 \frac{y(x^2 + y^2)^{\frac{1}{2}}}{(\frac{G}{f_1})} \frac{\sqrt{1-\varepsilon^2}}{f_2}$$

And

$$h_{2x} = -e_1 e_3 \frac{x(x^2+y^2)^{\frac{1}{2}}}{F}\sqrt{1-\varepsilon^2}$$

(4,13)

$$h_{2y} = e_2 \frac{y(x^2+y^2)^{\frac{1}{2}}}{(\frac{G}{f_1})}\frac{\varepsilon}{f_2}$$

Where $e_1, e_2, e_3 = \pm 1$ and $\varepsilon = \varepsilon(x,y)$ is an arbitrary functions of the coordinates $x, y$.

The integrability conditions for these two systems are simply:

$$h_{1xy} = h_{1yx} \quad \text{and} \quad h_{2xy} = h_{2yx} \tag{4.14}$$

Where $h_{1xy} = \frac{\partial^2 h_1}{\partial x \partial y}$ etc

The integrability conditions (4.14) permit to obtain another system of P.D.E's for the function $\varepsilon(x,y)$:

$$\frac{\varepsilon_x}{\sqrt{1-\varepsilon^2}} = -e_4 \frac{u}{v}\frac{x}{(x^2+y^2)}$$

(4.15)

$$\frac{\varepsilon_y}{\sqrt{1-\varepsilon^2}} = e_4 \frac{v}{u}\frac{y}{(x^2+y^2)}\left\{1-(x^2+y^2)\frac{1}{x}\frac{f_{2x}}{f}\right\}$$

Where

$$e_4 = \pm 1 \quad \text{and} \quad u = \frac{G}{f_1} \quad , \quad v = \frac{F}{f_2}$$

Again the integrability condition for this system is $\varepsilon_{xy} = \varepsilon_{yx}$ and this condition permits to prove by direct calculations the relations (4.9)-(4.11) and (4.8).

## 5. The case of Kerr metric

The functions F,G,E,H of Carter's class [A] of solutions, in the case of Kerr metric are given by relations (2.19):

$$G^2(y) = y^2 E^2(y) \qquad\qquad E^2(y) = y^2 - 2my + a^2 \tag{5.1}$$

$$F^2(x) = x^2 H^2(x) \qquad\qquad H^2(x) = a^2 - x^2 \tag{5.2}$$

A. Does the Kerr metric admit a congruence of ellipsoids of revolution?

The functions $h_1, h_2$ in this case, in the system of coordinates $x, y$ are given (see (3.10)-(3.12)) by:

$$h_1 = \frac{1}{a}\left(y^2 \pm l^2\right)^{\frac{1}{2}}\left(a^2 - x^2\right)^{\frac{1}{2}} \tag{5.3}$$

$$h_2 = \frac{1}{a}xy \tag{5.4}$$

If we substitute (5.1)-(5.4) in Theorem III we can deduce that:

1. Condition (4.5) is satisfied identically.

2. Condition (4.4) is satisfied if and only if:

$$\pm l^2 - a^2 = 0 \tag{5.5}$$

3. Condition (4.3) permit to find the function $f$ which describes the difference of the 3-dimensional metric of the ellipsoids in a curved space with that in a Euclidean space:

$$f^2 = \frac{y^2 - 2my + a^2}{y^2 + a^2} \tag{5.6}$$

Then we can state that the Kerr metric admit a class of observers defined by (4.1),(4.2) and with:

$$h_1 = \frac{1}{a}\left(y^2 + a^2\right)^{\frac{1}{2}}\left(a^2 - x^2\right)^{\frac{1}{2}} \tag{5.7}$$

The ellipsoids of revolution observed are necessarily oblate, if we permit prolate ellipsoids then from (5.5) the ellipsoids reduce to spheres and the Kerr metric reduces to the Schwarzschild solution! The equation of the ellipsoids of revolution in a Cartesian coordinates system is:

$$\frac{x_1^2 + x_2^2}{r^2 + a^2} + \frac{x_3^2}{r^2} = 1 \tag{5.8}$$

Where $y = r =$ constant, we see that the "ellipticity" of the surfaces is due to the angular momentum per unit mass of the Kerr metric in accordance with our "Classical Physical" intuition.

B. What is the most general form of surfaces of revolution admitted by the Kerr metric?

The answer to this question is a paper by its own but we can give a partial answer presenting the case of a surface of revolution admitted by the Kerr metric:

There are observers in the Kerr metric with quotient space that of a torus of revolution. The functions $h_1, h_2$ for a torus of revolution are given by (3.15), (3.16) in which we have applied the coordinate transformation (2.19c):

$$h_1 = R + \frac{1}{a}(y^2 \pm l^2)^{\frac{1}{2}}(a^2 - x^2)^{\frac{1}{2}} \tag{5.9}$$

$$h_2 = \frac{1}{a}xy \tag{5.10}$$

Obviously the relations (5.1)-(5.4) of Theorem III hold in the same way with the case of the ellipsoids of revolution, because the partial derivatives of $h_1, h_2$ are identical for the two cases due to the fact that R appeared in (5.9) is a constant.

Then we can state that the Kerr solution admits a class of observers defined by (4.1),(4.2) and with :

$$h_1 = R + \frac{1}{a}(y^2 + a^2)^{\frac{1}{2}}(a^2 - x^2)^{\frac{1}{2}} \tag{5.11}$$

and the observed congruence of surfaces is that of tori of revolution.

The tori of revolution are obtained by the revolution of an "oblate" ellipse around the axis of revolution, if we permit "prolate" ellipse to revolve about the axis of revolution then the ellipse reduces to a circle and the Kerr metric to the Schwarzschild one.

## 4. Concluding remarks

The definition of a foliation of the spaces of General Relativity by surfaces of revolution opens a great number of research directions.

1. Search for new vacuum stationary and axially symmetric solutions of Einstein's vacuum equations, taking as metric tensor that of relation (3.28).

2. Study of the existing interior solutions, candidates for the description of stars interior. Especially the study of the surface of the zero pressure of the Wahlquist solution ,as a surface of revolution can give an answer in the question ,why this solution can't be matched to Kerr metric.

3.Search of new interior solutions (describing perfect or anisotropic fluids) which could be matched to an exterior vacuum solution. Conditions (4.9)-(4.11) can be exploited to give an answer to this problem in the context of Carter's family of solutions.